\documentclass[11pt]{article}
\usepackage{amsmath,amssymb,color,graphics,epsfig,cite}

\usepackage{amsfonts}

\newcommand{\hoch}[1]{$\, ^{#1}$}

\newcommand{\be}{\begin{equation}}
\newcommand{\ee}{\end{equation}}
\newcommand{\bea}{\setlength\arraycolsep{2pt} \begin{eqnarray}}
\newcommand{\eea}{\end{eqnarray}}
\newcommand{\nn}{\nonumber}

\def\ft#1#2{{\textstyle{\frac{\scriptstyle #1}{\scriptstyle #2} } }}
\def\fft#1#2{{\frac{#1}{#2}}}

\def\0{{\sst{(0)}}}
\def\1{{\sst{(1)}}}
\def\2{{\sst{(2)}}}
\def\3{{\sst{(3)}}}
\def\4{{\sst{(4)}}}
\def\5{{\sst{(5)}}}
\def\6{{\sst{(6)}}}
\def\7{{\sst{(7)}}}
\def\8{{\sst{(8)}}}
\def\sst#1{{\scriptscriptstyle #1}}
\def\oneone{\rlap 1\mkern4mu{\rm l}}

\begin{document}

\begin{center}
{\Large {\bf Regular Black Holes and Stars from Analytic $f(F^2)$}}

\vspace{20pt}

Zhi-Chao Li\hoch{1} and H. L\"{u}\hoch{1,2}

\vspace{10pt}

{\it \hoch{1}Center for Joint Quantum Studies and Department of Physics,\\
School of Science, Tianjin University, Tianjin 300350, China }

\bigskip

{\it \hoch{2}Joint School of National University of Singapore and Tianjin University,\\
International Campus of Tianjin University, Binhai New City, Fuzhou 350207, China}
\vspace{40pt}

\underline{ABSTRACT}
\end{center}

We construct regular black holes and stars that are geodesically complete and satisfy the dominant energy condition from Einstein-$f(F^2)$ gravities with several classes of analytic $f(F^2)$ functions that can be viewed as perturbations to Maxwell's theory in weak field limit. We establish that regular black holes with special static metric ($g_{tt} g_{rr}=-1$) violate the strong energy condition and such a regular black hole with Minkowski core violates the null energy condition. We develop a formalism to perform electromagnetic duality transformations in $f(F^2)$. We obtain two new explicit examples where the duality is a symmetry. We study the properties of the corresponding dyonic black holes. We study the geodesic motions of a particular class of solutions that we call repulson stars or black holes.

\vfill{lizc@tju.edu.cn\ \ \  mrhonglu@gmail.com}


\thispagestyle{empty}
\pagebreak

\tableofcontents
\addtocontents{toc}{\protect\setcounter{tocdepth}{2}}

\newpage

\section{Introduction}

The Penrose singularity theorem, under the strong energy condition (SEC), relates the evolution of initial data to the spacetime structure \cite{Hawking:1973uf,Choquet-Bruhat:2009xil, Senovilla:2014gza}. If $R_{\alpha \beta} V^\alpha V^\beta \ge 0$ for all causal vectors and the initial data contain a closed trapped surface, the spacetime is geodesically incomplete. The nature of the incompleteness is not given, but it is typically believed to be due to the blow up of the curvature, although this is not proven. The situation is less clear for the inside of the event horizon of either static or stationary black holes, but many explicit regular black holes with no curvature singularity at the core have been constructed by relaxing the SEC. The best known example is the Bardeen metric \cite{bardeen}, which satisfies the weak energy condition (WEC). Regular black holes satisfying the dominant energy condition (DEC) are rarer, but known to exist in Einstein gravity coupled to quasi-topological electromagnetism \cite{Liu:2019rib}, with the coupling constant under some appropriate singular limit \cite{Cisterna:2020rkc}. Regular black holes are all fine-tuned objects, as there is always a free parameter associated with the condensation of gravitons in Einstein gravity that will generally create a curvature singularity at the core. There is recently a review on regular black holes that contains an extensive list of references \cite{Lan:2023cvz}.

Most of the constructions of regular black holes were {\it ad hoc}, writing up a desired metric before a theory, as in the case of the Bardeen black hole, which was later recognized as a nonlinear magnetic monopole \cite{Ayon-Beato:2000mjt}. An interesting technique was proposed \cite{Fan:2016hvf} to construct Einstein gravity coupled to some nonlinear electrodynamics (NLED) that admits spherically-symmetric and static black holes of the type\footnote{Two commentary notes on \cite{Fan:2016hvf} are worth mentioning \cite{Bronnikov:2017tnz, Toshmatov:2018cks}.}
\be
ds^2 = - h(r) dt^2 + \fft{dr^2}{h(r)} + r^2 (d\theta^2 + \sin^2\theta d\phi^2)\,,\label{ssm}
\ee
i.e. $g_{tt} g_{rr}=-1$. (See an related earlier work \cite{Bronnikov:2000vy}.) This was referred to as a special static metric in \cite{Li:2017ncu}. We shall adopt this terminology in this paper. The reverse engineering technique of \cite{Fan:2016hvf} shows that any such metric can be a magnetic monopole solution to Einstein-$f(F^2)$ gravity with certain specific $f(F^2)$, where $F^2=F^{\mu\nu} F_{\mu\nu}$ and $F_{\mu\nu}$ is the field strength of the Maxwell potential $A_\mu$. See a review on regular black holes sourced by nonlinear electrodynamics \cite{Bronnikov:2022ofk}.

However, the corresponding $f(F^2)$ theories for the Bardeen or Hayward \cite{Hayward:2005gi} metrics turn out to be disappointing since the functions are non-analytic with fractional powers acting directly on $F^2$. There is not much one can do with such a theory other than hanging on the wall. The statement is nevertheless significant and we can use it to prove two no-go theorems for a regular black hole with the special static metric: (1) it must violate the SEC; (2) it must violate the null energy condition (NEC) if it has the Minkowski core.

In this paper, we shall instead start with a sensible looking $f(F^2)$ and then construct regular black holes. It turns out that a regular black hole requires very little from $f(F^2)$. We must have $f(\infty)=0$ to ensure asymptotic flatness, and we must have finite $f(0)$ to ensure a regular core. These two conditions can be easily satisfied. We further require that $f(F^2)$ be an analytic function and can be viewed as a perturbation to Maxwell's theory under weak field limit, {\it i.e.}
\be
f(F^2) = -F^2 +\alpha_1 (F^2)^2 + \alpha_2 (F^2)^3 + \cdots\,.\label{wflimit}
\ee
Our construction simply assumes that the infinite sum gives rise to a closed-form expression that allows for the existence of regular black holes.  We construct several classes of such theories and obtain explicit regular black holes and stars satisfying the DEC, with Reissner-Nordstr\o m (RN) like asymptotic structure, extending a previously known example \cite{Kruglov:2014hpa,Kruglov:2021mfy}.

There are several advantages of this construction, in addition to having a sensible theory for further study. Since Maxwell's theory satisfies all the energy conditions, it is easier to construct its perturbation and hence also the closed-form expression that satisfies the DEC, the strongest energy condition possible after relaxing the SEC. Another important issue is that the de Sitter (dS) core of a regular black hole is not enough to ensure the geodesic completeness of the spacetime, and many popular ``regular'' black holes turn out to be geodesically incomplete \cite{Zhou:2022yio}. We show that the regular black holes constructed from our analytic $f(F^2)$ theories are necessarily geodesically complete.

These regular black holes are magnetically charged and for a purist, the ``regularity'' is still questionable since $F^2$ diverges at the $r=0$ core, even though the metric is perfectly regular. In this paper, we shall develop a formalism to perform electromagnetic duality transformation on $f(F^2)$ theory, and obtain the dual theory $\tilde f(G^2)$. The same regular black holes are now electrically charged. Furthermore, the electric field strength squared, {\it i.e.}~$G^2$, is finite all the way from the core to asymptotic infinity, removing the last objection to calling these spacetime configurations regular. Maxwell's theory has the electromagnetic duality symmetry. This symmetry is typically absent in general NLED. Born-Infeld theory \cite{Born:1933pep} is one rare example that retains the electromagnetic duality.  Our formalism allows us to construct $f(F^2)$ theories that possess the duality symmetry. We give two new explicit examples.

The paper is organized as follows.  In section 2, we consider Einstein-$f(F^2)$ gravity and show that charged spherically-symmetric and static black holes must have the special static metric. We obtain the equations of motion and review the reverse engineering technique of \cite{Fan:2016hvf}.  In section 3, we study the energy conditions and prove two no-go theorems for regular black holes with special static metrics. In section 4, we present several analytic $f(F^2)$ theories that admit explicit regular black hole or star solutions.  In section 5, we develop the formalism to perform electromagnetic duality on $f(F^2)$ theories, allowing us to obtain two explicit examples of $f(F^2)$ with the electromagnetic duality symmetry and study the properties of the dyonic black holes.  In section 6, we study the geodesic motion of regular black holes and stars.  We conclude the paper in section 7.

It is understood that this paper mainly deals with asymptotically-flat spacetimes, unless it is explicitly mentioned otherwise.

\section{The general setup}
\label{sec:setup}

We consider Einstein gravity minimally coupled to a class of NLED in the form of $f(F^2)$, where $F^2=F^{\mu\nu} F_{\mu\nu}$ and $F=dA$ is the field strength of the Maxwell potential $A$. The Lagrangian is
\be
{\cal L}=\sqrt{-g} \big(R + f(F^2)\big)\,.
\ee
The variation of the Maxwell potential $A_\mu$ gives
\be
\nabla_\mu (\varphi F^{\mu\nu})=0\,,\qquad \varphi(F^2)=f'(F^2)\,.\label{maxwell}
\ee
The Einstein equation associated with the variation of the metric $g_{\mu\nu}$ is
\be
G_{\mu\nu} \equiv R_{\mu\nu} - \ft12 g_{\mu\nu} R=\ft12 \Big(-4\varphi F^2_{\mu\nu} + g_{\mu\nu} f(F^2)\Big)\,.
\ee

\subsection{Special static metrics}

In four dimensions, the spherically-symmetric and static solutions can carry both electric and magnetic charges. The most general ansatz takes the form
\bea
ds^2 &=& - h(r) e^{-\sigma(r)} dt^2 + \fft{dr^2}{h(r)} + r^2 \big(d\theta^2 + \sin^2\theta d\phi^2\big)\,,\cr
F&=&\psi(r) dt\wedge dr + p \sin\theta d\theta \wedge d\phi\,,\label{genans}
\eea
where $\psi$ is the radial electric field and $p$ is the magnetic charge parameter.
(See an related earlier work \cite{Bronnikov:2000vy}.) It is easy to verify that
\be
-G_t{}^t + G_r{}^r = -\fft{h}{r} \sigma'=0\,.
\ee
We therefore set $\sigma=0$, without loss of generality. The resulting metric, taking the form of \eqref{ssm}, is called a special static metric in \cite{Li:2017ncu}. This implies that
\be
F^2=-2\psi^2 + \fft{2p^2}{r^4}.
\ee
Consequently, we have
\be
G_{t}{}^t + G_{r}{}^r = \fft{2(rh' + h -1)}{r^2} = f(F^2) + 4\varphi \psi^2.\label{eq1}
\ee
With $\sigma=0$, the Maxwell equation can be easily solved, independent of the metric function $h$, giving
\be
\psi = \fft{q}{\varphi r^2}\,,\qquad \hbox{or}\qquad \varphi=0\,.
\ee
Note that the existence of the second case $\varphi=0$ depends on the structure of the theory. For example, in Maxwell's theory, $f(F^2)=-F^2$, $\varphi=-1$. For general discussions in this section, we therefore consider only the first case, where $q$ is the parameter of electric charges. (We relegate the discussion of $\varphi=0$ to section \ref{sec:constante}.) It is now easy to verify that the remaining Einstein equations in the sphere direction are automatically satisfied, which amounts to acting on \eqref{eq1} with $(r\partial_r + 2)$.

\subsection{Reverse engineering technique and its critique}

Typically one constructs black holes from a given specific theory. Reverse engineering technique is a method to determine the appropriate theory for a given metric. This technique has been used to construct black holes in Einstein-scalar theory, see e.g.~\cite{Anabalon:2012ta, Anabalon:2013qua,Feng:2013tza}. A potential danger is that the black hole in question is not the most general solution to the constructed theory. In many cases, the black hole parameters, even including the mass, may become coupling constants of the theory.

The reverse engineering technique of \cite{Fan:2016hvf} is particularly simple in $f(F^2)$ theory, since the equations of motion effectively reduce to the first-order equation \eqref{eq1}. Although we can in principle perform electric and magnetic duality in a general $f(F^2)$ theory, it is actually rather subtle and we shall discuss this in section \ref{sec:emdual}. It is thus more convenient to discuss the reverse engineering technique in the pure magnetic case, for which the complete set of equations of motion reduces to \eqref{eq1} with $\psi=0$. This implies that for any metric function $h(r)$, we can obtain $f(F^2)$ as a function of $r$, thus for magnetic solutions, one can determine the $f(F^2)$ by
\be
f(F^2)= \fft{2(rh' + h -1)}{r^2}\Big|_{r\rightarrow (\fft{2p^2}{F^2})^{\fft14}}.
\ee
This leads to \cite{Fan:2016hvf}
\bea
\hbox{Bardeen}:&& h = 1 - \fft{2M r^2}{(r^2 + g^2)^{\fft32}}\,,\qquad\rightarrow\qquad f=\fft{1}{\alpha}\fft{(\alpha F^2)^{\fft54}}{\big(1 + \sqrt{\alpha F^2}\big)^{\fft52}}\,,\\
\hbox{Hayward}:&& h = 1 - \fft{2M r^2}{r^3 + g^3}\,,\qquad\rightarrow\qquad f=\fft{1}{\alpha}\fft{(\alpha F^2)^{\fft32}}{\big(1+(\alpha F^2)^{\fft34}\big)^2}\,.
\eea
(The NLED theory for the Bardeen metric was obtained earlier \cite{Ayon-Beato:2000mjt}.) It is certainly disappointing to see that fractional powers, arising mainly from the reversing relation $r\sim 1/(F^2)^{\fft14}$, act directly on $F^2$. Although there is no branch cut in these specific magnetic solutions, the general theory surely does. Frankly there is not much one can do about these theories, except for constructing these specific solutions.

Nevertheless, there is an important lesson one can learn from the fact that there exists an $f(F^2)$ for any special static metric. This helps us to establish some no-go theorems of regular metrics, which we shall discuss next. We shall resume the constructions of regular black holes in section \ref{sec:explicit}.

\section{Energy conditions and no-go theorems}
\label{sec:no-go}

For spherically-symmetric and static black holes, the metric is diagonal and so is the energy-momentum tensor $T^{\mu}_{\nu} = {\rm diag}\{-\rho,p_r,p_T,p_T\}$. Owing to the Einstein equation, we can express the energy-momentum tensor using either the metric fields or the matter fields. For the special static metrics, we have
\bea
\rho=-p_r = -\fft{h'}{r} - \fft{h-1}{r^2}= -\ft12 (f + 4\varphi \psi^2)\,,\qquad
p_T=\ft12 h'' + \fft{h'}{r}=\ft12 f - \fft{2\varphi p^2}{r^4}\,.
\eea
Owing to the fact that $\rho+p_r=0$ for special static metrics \eqref{ssm}, the energy conditions for $h>0$ and $h<0$ take the same form:
\bea
\hbox{NEC}:&& \rho+p_T\ge 0\,;\qquad \hbox{WEC}:\quad \rho\ge 0\oplus \hbox{NEC}\,;\qquad
\hbox{DEC}:\quad \rho-p_T\ge 0 \oplus \hbox{WEC}\,;\nn\\
\hbox{SEC}:&& p_T\ge 0 \oplus \hbox{NEC}\,.
\eea
This implies that when we discuss the special static metrics, we can simply examine the energy and pressure density in the whole $r\in (-\infty, \infty)$ region, ignoring the subtlety of whether the metric has a horizon or not.

\subsection{Energy conditions on a regular core}

In the next section, we shall construct some explicit regular black holes or stars from Einstein-$f(F^2)$, satisfying the DEC, but not the SEC. This raises the question of whether the SEC prohibits such spacetime structures. For the spherically-symmetric and static spacetime to be absent from a singularity in the middle, it could be a wormhole \cite{Ellis:1973yv}, black bounce \cite{Simpson:2018tsi} or a dark wormhole that connects to dS universe \cite{Geng:2015kvs}. In Einstein gravity with minimally coupled matter, they will inevitably violate the NEC. A simpler choice is that the $r\rightarrow 0$ core asymptotically approaches Minkowski, dS or anti-de Sitter (AdS) spacetimes where the metric function behaves as
\be
h = 1 + a_2 r^2 + a_3 r^3 + {\cal O}(r^3)\,,\qquad
h e^{-\sigma} = b_0 + b_2 r^2 + b_3 r^3 + {\cal O}(r^3)\,,\qquad b_0>0\,.
\ee
The SEC and DEC conditions imply that
\be
\hbox{SEC}:\quad b_2> {\rm max}\{0, a_2b_0\}\,;\qquad
\hbox{DEC}:\quad a_2< 0\,,\qquad a_2b_0 < b_2< -2a_2 b_0\,.
\ee
This star-like core configuration can satisfy both the DEC and SEC simultaneously. This is expected since we certainly believe that the interior matter of our Sun (ignoring its rotation) satisfies both conditions.

However, the regular black holes or stars from Einstein-$f(F^2)$ gravity have an additional constraint, namely $\sigma=0$, implying that $b_0=1$, $a_2=b_2$, $a_3=b_3$, {\it etc}.
The characteristics of the core can be determined by the curvature, namely
\be
a_2<0\ \ \rightarrow\ \  \hbox{dS core};\qquad a_2=0 \ \ \rightarrow\ \
\hbox{Mink core};\qquad a_2>0\ \ \rightarrow \ \ \hbox{AdS core}.
\ee
The WEC or DEC requires a dS core with $a_2<0$, but the SEC requires an AdS core with $a_2>0$.
If the core is Minkowski with $a_2=0$, the same argument proceeds with $a_3$, and so on.
Thus, we can conclude that even with the $g_{tt} g_{rr}=-1$ constraint, a regular core does not necessarily violate either DEC or SEC, but it does violate WEC $\oplus$ SEC.

\subsection{No-go theorem of regular black holes under SEC}

Although the regular core does not violate the SEC locally, further imposing asymptotic flatness at large $r$ will, at least for the special static metrics. This is based on the earlier proof that any such nontrivial metric can be a magnetic monopole of a certain Einstein-$f(F^2)$ theory with
\be
\rho =-p_r= - \ft12 f(\chi)\,,\qquad p_T = \ft12 f(\chi) - \chi f'(\chi)\,,\qquad
\chi= \fft{2p^2}{r^4}>0\,.
\ee
The SEC requires
\be
\rho+p_T = -\chi f'(\chi)\ge 0\,,\qquad p_T\ge 0\,.
\ee
Note that the first condition above is also the NEC condition in the transverse direction. It should be kept in mind that $r\rightarrow 0$ corresponds to $\chi\rightarrow \infty$ and $r\rightarrow \infty$ corresponds to $\chi=0$.  Therefore, the asymptotic flatness at large $r$ requires $f(0)=0$ and regularity at $r=0$ requires that $f(\infty)$ be a constant. In this case $\chi f'(\chi)$ must vanish as $\chi\rightarrow \infty$. Since $f'(\chi)$ must be negative due to the NEC, $f(\infty)$ must be a negative constant, indicating that the core is dS-like.  Thus, the strong energy condition must be violated at the core.

Earlier, we established that the local regular core does not necessarily violate the SEC, even for the special static metrics.  However, now we see that imposing the asymptotic flatness at large distances and NEC everywhere, regular special static metrics necessarily violate the SEC in the core. The violation may not be surprising since the regular special static metrics imply that the time ticking rates for a static object are the same at both asymptotic infinity and the core, which does not conform to the expectation that time dilation should have occurred. Therefore, this no-go theorem on its own does not have much to say about more general static metrics where $g_{tt} g_{rr}$ is not constant.

\subsection{No-go theorem of regular black holes with Minkowski core}

In order to have a Minkowski core, we must have $f(\infty)=0$. The asymptotic flatness at large $r$ requires that $f(0)=0$. This implies that for non-vanishing $f$, there must be a minimum of $f(\chi)$ and hence we must also have $f'(\chi)>0$ at a certain spacetime region. This immediately leads to a violation of the NEC in the transverse direction.

Before finishing this section, we would like to emphasize that although we made use of the tool of the reverse engineering technique, our two no-go theorems apply to all regular black holes with special static metrics.

\section{Regular black holes and stars from analytic $f(F^2)$}
\label{sec:explicit}

Having established the no-go theorem under SEC, we shall be content to construct regular black holes satisfying DEC. For a simple equation such as \eqref{eq1}, we do not really need the reverse engineering technique since the solution is already a quadrature:
\be
h= 1- \fft{2\mu}{r} + \fft{1}{2r}\int_0^r \tilde r^2 f(F^2) d\tilde r\,,\qquad F^2 = \fft{2p^2}{\tilde r^4}\,.\label{gensol}
\ee
The absence of singularity at $r=0$, corresponding to $F^2\rightarrow \infty$, requires that
\be
\lim_{F^2\rightarrow \infty}  f(F^2)={\rm constant}\,.\label{cond1}
\ee
For the solution to be (AdS)-RN-like asymptotically, we may also require that
\be
\lim_{F^2\rightarrow 0} f(F^2)\sim  -2\Lambda_0 - F^2.\label{conv1}
\ee

\subsection{A simple example}
\label{sec:simple}

For the most part of our paper, we shall set the bare cosmological constant $\Lambda_0=0$. Such examples with (\ref{cond1},\ref{conv1}) can be easily constructed, and we shall present several classes of such theories in section \ref{sec:further}. Their magnetically-charged solutions have analogous properties and we select the simplest example to illustrate the general feature here. The simplest case perhaps is
\be
f=-\fft{F^2}{1+\alpha F^2}\,,\qquad \alpha\ge 0\,.\label{simplest}
\ee
The theory was first proposed in \cite{Kruglov:2014hpa}. The theory reduces to the standard Maxwell theory when $\alpha=0$. For small $\alpha$, this can be viewed as a certain order-by-order higher-derivative correction to Maxwell's theory, which just happens to be able to be summed over to a closed-form expression.

For the magnetic ansatz, the metric function $h$ is
\be
h=1 - \fft{2M}{r} + \fft{p^2}{r^2}\, {}_2F_1[\ft14,1;\ft54; -\fft{2\alpha p^2}{r^4}]\,.\label{oursol1}
\ee
The hypergeometric function can also be expressed in terms of simpler special functions
\bea
{}_2F_1[\ft14,1;\ft54; -\fft{1}{x}]&=&\fft{\sqrt[4]{x}}{4\sqrt2}\Bigg(2(\pi + \arctan(1-\sqrt2 \sqrt[4]{x}) -\arctan(1+\sqrt2 \sqrt[4]{x})\cr
&& -\log\Big(\frac{-\sqrt{2} \sqrt[4]{x}+\sqrt{x}+1}{\sqrt{2} \sqrt[4]{x}+\sqrt{x}+1}\Big)
\Bigg).\label{hypergeom}
\eea
The metric with these special functions was given in \cite{Kruglov:2021mfy} where the black hole properties were analyzed. It is straightforward to verify that the solution satisfies the DEC for all real $r$.

The metric is asymptotic to Minkowski spacetime at large $r$, and its leading falloffs are RN-like carrying the magnetic charge, with
\be
h= 1 - \fft{2M}{r} + \fft{p^2}{r^2} -\frac{2 \alpha  p^4}{5 r^6}+ {\cal O}(r^{-10})\,.
\ee
For sufficiently large $M$, there are roots of function $h$, the event horizon is the largest root $r_+$. It is straightforward to obtain the complete set of the thermodynamic variables
\bea
&&M=\frac{r_+}{2} + \frac{p^2}{2 r_+}\, _2F_1\left(\frac{1}{4},1;\frac{5}{4};-\frac{2\alpha p^2 }{r_+^4}\right)\,,\quad T=\frac{1}{4 \pi  r_+}-\frac{p^2 r_+}{4 \pi  \left(2 \alpha  p^2+r_+^4\right)}\,,\quad S=\pi r_+^2\,,\nn\\
&&Q_m=p\,,\qquad \Phi_m=\frac{3 p}{4 r_+} \, _2F_1\left(\frac{1}{4},1;\frac{5}{4};-\frac{2\alpha p^2}{r_+^4}\right)+\frac{p r_+^3}{8 \alpha  p^2+4 r_+^4}\,,
\eea
and verify the first law $dM=TdS + \Phi_m dQ_m$. For our convention \eqref{conv1}, the magnetic charge and its potential are defined by
\be
Q_m = \fft{1}{4\pi} \int F\,,\qquad
\Phi_m = -\int_{r_+}^\infty dr\, \fft{p}{r^2} \varphi\,.
\ee
Since the solution satisfies the DEC and trace energy condition (TEC), it follows that it satisfies both the upper and lower bounds on temperature \cite{Khodabakhshi:2022jot}
\be
\sqrt{\fft{1}{\pi\, S}} - \fft{M}{S}\, \le 2T \le\, \fft{3M}{S} - \sqrt{\fft{1}{\pi S}}\,.\nn
\ee
Note that the upper bound requires only the NEC and was established in \cite{Yang:2022yye}.

\begin{figure}[ht!]
\begin{center}
\includegraphics[width=220pt]{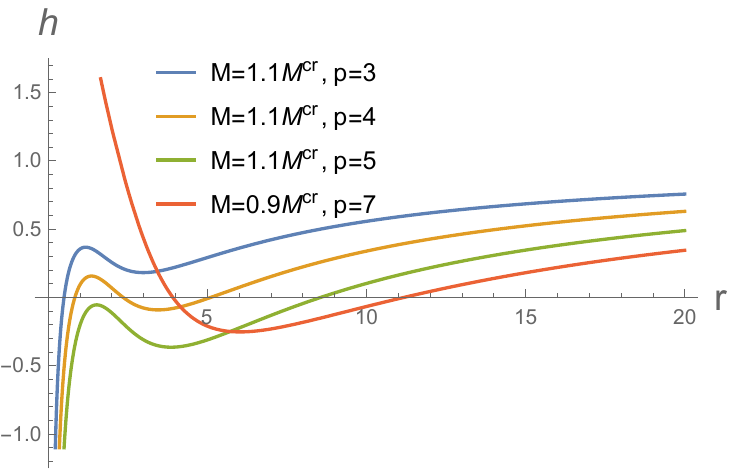}\ \
\includegraphics[width=220pt]{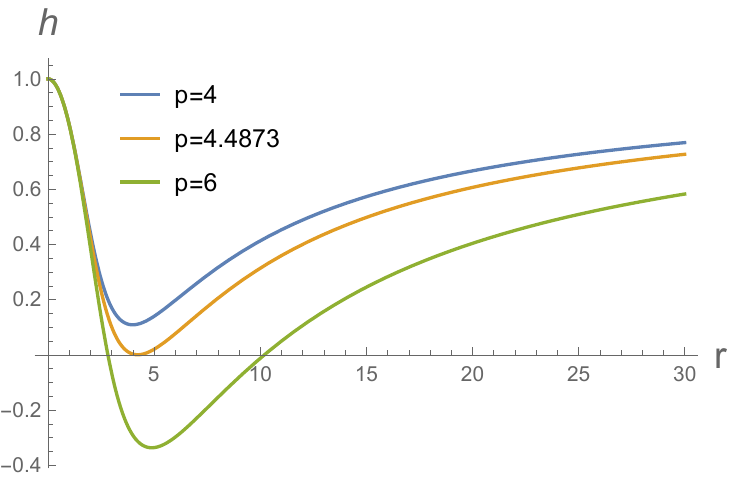}
\end{center}
\caption{{\it\small The metric function $h$ of singular and regular solutions. The left panel displays singular black holes with $M\ne M^{\rm cr}$. The singularity is time-like when $M>M^{\rm cr}$, with either one or three horizons. It is space-like when $M<M^{\rm cr}$ with two horizons. The right panel shows regular solutions with $M=M^{\rm cr}$, associated with repulson star, extremal and nonextremal black holes. In both panels, we have set $\alpha=1$.}}
\label{metrich}
\end{figure}

The $f(F^2)$ theory \eqref{simplest} was motivated to construct regular black holes. Regular special static metrics are all similar, and well studied in literature. For later purpose and also completeness, we shall give a quick review here. Indeed, at $r=0$, where curvature singularity can arise, we have
\be
h = 1 - \fft{2(M-M^{\rm cr})}{r} - \fft{r^2}{6\alpha} + {\cal O}(r^3)\,,\qquad
M^{\rm cr} = \fft{\pi p^{\fft32}}{4 (8 \alpha)^{\fft14}}\,.
\ee
The solution is absent of curvature singularity provided that $M=M^{\rm cr}$. It should be pointed out that this is a fine-tuned result since a slight increase or decrease of the mass away from $M^{\rm cr}$ will yield singularity. For $M> M^{\rm cr}$, the singularity at $r=0$ is time-like, analogous to the Schwarzschild black hole. When $M<M^{\rm cr}$, the singularity is space-like, analogous to the RN black hole, as can be seen in the left panel of Fig.~\ref{metrich}. (For $M<M^{\rm cr}$, $p$ must be sufficiently large to avoid naked singularity.) Note that the number of black hole horizons (including the inner ones) can be 1, 2 and 3. It was shown that black holes satisfying SEC can have at most two horizons \cite{Yang:2021civ}, but such a restriction does not exist for black holes under DEC. Black holes satisfying DEC with four horizons were constructed in Einstein gravity coupled to quasi-topological electromagnetism \cite{Liu:2019rib}.

When $M=M^{\rm cr}$, the metric is regular and geodesically complete, and three spacetime structures can arise. When $p=p^*\equiv 4.4873\sqrt{\alpha}$, corresponding to
\be
M^{\rm cr}_{\rm ext}=4.43909 \sqrt{\alpha}\,,\qquad
r_+=r_-=4.22895 \sqrt{\alpha}\,.
\ee
The solution is a regular extremal black hole. When $p>p*$, the regular black hole becomes nonextremal with two horizons. When $p<p^*$, the solution describes a star with no horizons.
The fact that the function $h$ has a minimum implies that the static gravity force is repulsive close to the core. We shall refer to this spacetime configuration as a ``repulson star.''

Regardless of the number of the horizons, the general first law established earlier holds for all these special cases. However, the situation needs a special mention for the repulson star since we no longer have a horizon. In this case, the first law of repulson star dynamics is
\be
dM^{\rm cr} = \Phi_m dQ_m\,,\qquad \Phi_m = -\int_{0}^\infty dr\, \fft{p}{r^2} \varphi\,.
\label{fostar}
\ee
An analogous first law was also obtained for electrically-charged boson stars \cite{Liu:2020uaz}.

It is worth pointing out that although the metric is regular, the field strength $F^2$ is divergent at the core due to the construction of the magnetic monopole. However, the finite $\Phi_m$ of the repulson star \eqref{fostar}, which is in fact the potential of the electric dual of $F$, strongly suggests that the dual electrically-charged solution would be regular not only in the metrics, but also in the field strength. We shall come back to discuss this in section \ref{sec:emdual}.

\subsection{Geodesic completeness}

When $M=M^{\rm cr}$, the curvature is regular at the core $r=0$ with a positive cosmological constant. However, having a dS core does not guarantee that the spacetime is geodesically complete at $r=0$, if the function $h$ is not an even function of radial coordinate $r$ \cite{Zhou:2022yio}. In our case, the regular metric is indeed even on $r$.  This can be seen more explicitly in the small $r$ expansion:
\be
h(r)= 1-\frac{r^2}{6 \alpha }+\frac{r^6}{28 \alpha ^2 p^2}-\frac{r^{10}}{88 \left(\alpha ^3 p^4\right)}+\frac{r^{14}}{240 \alpha ^4 p^6}-\frac{r^{18}}{608 \left(\alpha ^5 p^8\right)}+{\cal O}\left(r^{22}\right).
\ee
A careful examination of the hypergeometric function \eqref{hypergeom} can determine that
the regular $h(r)$ of \eqref{oursol1} is indeed an even function of $r$. In fact, the evenness of the function $h$ was guaranteed in our construction, which insists that the function $f(F^2)$ is an analytic function of $F^2$ that is infinitely differentiable. Since $F^2=2p^2/r^4$, $f(F^2)$ must be an even function of $r$, and hence is the $h(r)$ given by \eqref{gensol} for $\mu=0$.

Since the metric is an even function of $r$, the regions of negative $r$ are identical to those of positive $r$, and hence there is no curvature singularity everywhere in $r\in (-\infty,\infty)$. Since the foliating 2-sphere collapses at $r=0$, we can have an option of identifying the negative and positive regions so that $r$ runs only from 0 to plus infinity.
This argument applies to all our regular examples constructed in this paper.

\subsection{Further examples}
\label{sec:further}

The simple $f(F^2)$ theory \eqref{simplest} is one of many such examples one can construct. We require that the theory has one coupling constant $\alpha$ and for small $\alpha$, $f(F^2)$'s are perturbative higher-order corrections to Maxwell's theory.  We further require that the metric can be analytically solved and contains regular black holes and stars. Here are several further examples. We present both the theory and metric function $h$ of the magnetic solution.
We shall not discuss their global structure since, as mentioned earlier, the essential characteristics are the same as that of the earlier simple case.

The first class of $f(F^2)$ theories is given by
\bea
f &=&\fft{1}{\nu\alpha}\Big(\fft{1}{(1 + \alpha F^2)^\nu}-1\Big)\,,\nn\\
h &=& 1- \fft{2M}{r} + \fft{r^2}{6\nu\alpha}
\Big({}_2F_1[-\ft34,\nu;\ft14;-\alpha\ft{2p^2}{r^4}]-1\Big)\,.
\eea
The $\nu=1$ case yields the earlier simple example.  This series contains Maxwell's theory.
Specifically, when $\nu=-1$ and $-1/2$, the theory is the standard Maxwell theory or BI theory (of purely electric or magnetic charges) respectively.
For $\nu<-1$, the theories violate the NEC; for $-1\le \nu\le -1/2$, they satisfy both the DEC and SEC; for $\nu >-1/2$, they satisfy DEC, but not SEC, and regular magnetic monopoles emerge.

The second class of theories takes the form
\bea
f &=& -\fft{F^2}{(1 + \alpha (F^2)^\mu)^\nu}\,,\nn\\
h&=& 1 - \fft{2M}{r} + \fft{p^2}{r^2}{}_2F_1[\ft{1}{4\mu},\nu; 1+ \ft{1}{4\mu};
-\alpha (\ft{2p^2}{r^4})^\mu]\,.\label{moreexample1}
\eea
Suitable energy conditions require that $\mu\nu\le 1$, in which case, DEC is satisfied. In all our regular solutions satisfying the DEC, the cores at $r\rightarrow 0$ are necessarily dS. In the above example, if $\mu\nu>1$, $f(\infty)=0$, and hence the solution will have the Minkowski core. However, as was established, these solutions necessarily violate the NEC.

The third example is given by
\bea
f &=& \fft{1}{2\alpha} (e^{-2\alpha F^2}-1)\sim -F^2 + \alpha (F^2)^2 - \fft23\alpha^2 (F^2)^3 + \cdots\,,\nn\\
h&=& 1 - \fft{2\mu}{r} + \fft{r^2}{48\alpha}\Big(E_{\fft74}(\fft{4\alpha p^2}{r^4})-4
\Big)\,.
\eea
It can be easily verified that the general solution satisfies DEC provided that $\alpha>0$.

Finally we consider
\be
f=\fft{1}{2\alpha} (e^{-2\alpha F^2}-1) e^{-2\alpha F^2}\,.
\ee
The function vanishes at both $F^2\rightarrow \infty$ and $F^2\rightarrow 0$. It gives rise to a regular black hole with the Minkowski core
\be
h(r\rightarrow 0)\sim 1-\frac{r^6 e^{-\frac{4 \alpha  p^2}{r^4}}}{64 \alpha ^2 p^2} + \cdots\,.
\ee
As we have proven generally, this regular black hole violates NEC. It is worth emphasizing again before finishing this section that all our regular black holes are geodesically complete.

\section{Electromagnetic duality via scalar-vector theories}
\label{sec:emdual}

\subsection{Electromagnetic duality transformation}

In the previous section, we constructed several classes of explicit examples of $f(F^2)$ theories that admit regular black holes or stars. All of these solutions carry magnetic charges. A direct construction requiring both analyticity of $f(F^2)$ and regularity of black holes turns out to be tough for electrically-charged configurations, and there were no-go theorems against regular black holes carrying electric charges \cite{Bronnikov:2000vy, Bokulic:2022cyk}. However, Maxwell's equation of motion \eqref{maxwell} implies that there exists a dual version of the $f(F^2)$ theory where the same regular metrics are instead supported by the electric charges.

The electromagnetic duality transformation follows directly from Maxwell's equation \eqref{maxwell}, given by
\be
\varphi\, {*F} = G\equiv dB,\label{emdual0}
\ee
where $B$ is the potential of the dual field strength $G$. This equation implies that
\be
\varphi^2 F^2 = - G^2.\label{emdual1}
\ee
We can thus solve for $F^2$ in terms of $G^2$. However, the minus sign illustrates a subtlety of electromagnetic duality. For Maxwell's theory with $\varphi=-1$, the minus sign shows that substituting the dual relation above into the Lagrangian will yield a wrong sign of the dual theory. In the case of Maxwell's theory, the dualization can be done properly in the first-order formalism, namely
\be
{\cal L} = f(F^2) {*\oneone} + 2 dB\wedge F\,.
\ee
The variation with respect to $B$ yields $dF=0$, giving rise to the $f(F^2)$ theory. Variation of $F$ gives us duality equation \eqref{emdual0}. However, since for general $f(F^2)$, we do not know how to express $F$ in terms of $G$ explicitly, we do not have a way of evaluating the $dB\wedge F$ term, as a function of $G$.  This subtlety leads us to consider the equivalent scalar-vector theory of $f(F^2)$.

\subsection{Scalar-vector and $f(F^2)$ theories}
As in the case of $f(R)$ gravity, we can express an $f(F^2)$ theory in terms of an equivalent scalar-vector theory, namely
\be
{\cal L}=\sqrt{-g} (-\phi^2 F^2 - V(\phi))\,,\label{scalarvector}
\ee
where
\be
\phi^2 = -f'(\chi)\,,\qquad V=\chi f'(\chi) -f(\chi)\,.
\ee
It should be pointed out that the inverse function of $f'(\chi)$ may not always have a closed-form expression, in which case, the potential $V$ in \eqref{scalarvector} cannot be expressed by $F^2$ in a closed form. For example, this procedure does not work well for the $f(F^2)$ theories associated with the Bardeen or Hayward black holes.

Our examples of $f(F^2)$ are simple analytical functions and $f'(\chi)$'s have their inverses. For example, it is straightforward to obtain the scalar-vector theory for the simplest example:
\be
f=-\fft{F^2}{1 + \alpha F^2}\,,\qquad \leftrightarrow \qquad V = \fft{1}{\alpha} (\phi-1)^2\,.
\label{simplestscalar}
\ee
Here are the further examples of our $f(F^2)$ theories and their scalar-vector equivalences:
\bea
f=-\fft{F^2}{(1 + (\alpha F^2)^n)^{\fft1{n}}}\,,\qquad&\leftrightarrow&\qquad
V=\fft{1}{\alpha} \left(1-\phi ^{\frac{2 n}{n+1}}\right)^{\frac{n+1}{n}},\nn\\
f= \fft{1}{\nu \alpha} \Big(\fft1{(1 + \alpha F^2)^\nu} - 1\Big)\,,\qquad
&\leftrightarrow&\qquad  V=\fft{1}{\alpha \nu} \Big(1+\nu \phi^2-(\nu+1) \phi^{\frac{2\nu}{\nu+1}}\Big)\,,\nn\\
f= \fft{1}{2\alpha} (e^{-2\alpha F^2}-1)\,,\qquad&\leftrightarrow&\qquad
V=\fft1{2\alpha} (1-\phi^2 + \phi^2 \log\phi^2)\,.
\eea
Note that although fractional powers on a kinetic term such as $F^2$ are abhorrent, they are not always problematic in the scalar potential.

Having obtained the scalar-vector theory \eqref{scalarvector} for a relevant $f(F^2)$ theory, it is straightforward to obtain its electromagnetic dual, given by
\be
{\cal L} = \sqrt{-g} \big(-\phi^{-2} G^2 - V(\phi)\big)\,.\label{dualtheory}
\ee
This allows us to derive the electromagnetic duality perturbatively, namely
\be
{\cal L}=\sqrt{-g} \Big(-F^2 + \alpha (F^2)^2 + \beta (F^2)^3 + \cdots\Big)\,,\label{emdual2}
\ee
is dual to
\be
{\cal L} = \sqrt{-g} \Big(- G^2 + \alpha (G^2)^2 - (\beta + 4\alpha^2) (G^2)^3 + \cdots\Big)\,.
\label{emdual3}
\ee
We thus see that up to and including the subleading order, the electromagnetic duality of Maxwell's theory remains a symmetry. The duality continues to be the symmetry at the $(F^2)^3$ order provided that $\beta=-2\alpha^2$.

As a concrete example, we now perform the electromagnetic duality transformation on the simplest NLED \eqref{simplest}. The scalar-vector theory of the dual theory is given by \eqref{dualtheory} with the scalar potential given by \eqref{simplestscalar}. The scalar equation of motion implies that
\be
\phi ^4-\phi ^3=-\alpha  G^2.
\ee
Substituting the appropriate $\phi$ as a function of $G^2$ back to \eqref{dualtheory} yields the dual theory $\tilde f(G^2)$. The closed-form expression of the quartic-order polynomial exists, but is somewhat unpresentable. We give both the small and large $G$ limits:
\bea
\tilde f(G^2) &=& - G^2 + \alpha (G^2)^2 -3 \alpha^2 (G^2)^3 + {\cal O}\big((G^2)^4\big)\,,\nn\\
\tilde f(G^2) &=& -\sqrt{\fft{4G^2}{\alpha}} + \left(\fft{16 G^2}{\alpha^3}\right)^{\fft14} +
\fft{3}{4\alpha} + {\cal O}\big((G^2)^{-\fft14}\big)\,,
\eea
which confirm the duality relation established order by order between \eqref{emdual2} and \eqref{emdual3} in the weak-field limit. This is in fact the general feature of our analytic $f(F^2)$ theory. The duality transformation \eqref{emdual1} implies that $G^2$ as a function of $F^2$ is infinitely differentiable for our requirements of both analyticity and weak-field limit \eqref{wflimit}, and hence vice versa, $F^2$ as a function of $G^2$ is also infinitely differentiable. This ensures that our regular magnetic monopole metrics are also the well-defined electrically-charged solutions of the dual theory.

However, for the electrically-charged solution, we have $G^2=-2\psi^2$. It would thus be problematic if the electric field $\psi$ becomes large at the regular core, since $f(G^2)$ is not analytic at large $G^2$. However, the duality transformation \eqref{emdual1} in this case become
\be
G^2 = -\fft{F^2}{(1 + \alpha F^2)^4}\,,
\ee
which vanishes at both asymptotic infinity $r\rightarrow \infty$ and the regular core $r=0$, since $F^2=2p^2/r^4$. Therefore, although $F^2$ diverges at the core in the regular magnetic monopole, $G^2$ of the dual theory is finite from the core to asymptotic infinity, giving rise to an electrically-charged black hole or star that is regular not only in the metric, but also in the field strength.

\subsection{$f(F^2)$ theories with electromagnetic duality symmetry}

Electromagnetic duality is a symmetry of Maxwell's theory $f=-F^2$. As we see earlier, it is no longer a symmetry for general $f(F^2)$ NLED theories. However, if it can be expressed as
the scalar-vector theory \eqref{scalarvector} with $V(\phi)=V(\phi^{-1})$, the duality transformation remains a symmetry. The simplest such example is
\be
V=\fft{1}{\alpha} (\phi^2 + \phi^{-2} -2)\,,
\ee
corresponding to
\be
f=\fft{2}{\alpha} \Big(1 - \sqrt{1+\alpha F^2})\,.\label{emdualf1}
\ee
This theory is effectively the BI theory for purely electric or purely magnetic cases, but it is not the BI field since it lacks the term $(\epsilon^{\mu\nu\rho\sigma} F_{\mu\nu} F_{\nu\rho})^2$. Nevertheless, the theory has the electromagnetic duality symmetry, as in the case of the BI theory, despite the absence of the term.

By the same logic, we can construct more examples of $f(F^2)$ that have electromagnetic duality. We present two explicit examples. Consider the potential
\be
V=\fft{1}{2\alpha}-\fft{1}{\alpha (\phi^2 + \phi^{-2})}\,,
\ee
it leads to
\be
{\cal L}=\fft{1}{8\alpha}\left(\sqrt{1+\sqrt{1+8 \alpha  F^2}} \left(3-\sqrt{1+8 \alpha  F^2}\right)^{3/2}-4\right).\label{emdualf2}
\ee
Another example is provided with
\be
V=\fft{12}{\alpha} (\phi^{\fft23} + \phi^{-\fft23}-2)\,,
\ee
which leads to
\be
{\cal L} = \fft{24}{\alpha} - \fft{8\sqrt2 (2 + \sqrt{1 + \alpha F^2})}{\alpha
\sqrt{1 + \sqrt{1 + \alpha F^2}}}\,.\label{emdualf3}
\ee
It can be checked that for small $F^2$ expansion, both of our new examples have the same
structure of \eqref{emdual2} with $\beta=-2\alpha^2$, up to and including the $(F^2)^3$ order, with the understanding that appropriate scaling of $\alpha$ may be needed.

\subsection{Dyonic black hole}

It can be argued that the $f(F^2)$ of \eqref{emdualf1} is the simplest NLED that preserves electromagnetic duality. The dyonic black hole is given by \cite{Bokulic:2022cyk}
\bea
h&=& 1-\fft{2M}{r}+\frac{r^2}{3 \alpha } \left(1-F_1\left(-\frac{3}{4};-\frac{1}{2},-\frac{1}{2};\frac{1}{4};-\frac{2 p^2 \alpha }{r^4},-\frac{2 q^2 \alpha }{r^4}\right)\right),\cr
A &=&\frac{q \sqrt{2 \alpha  p^2+r^4}}{r^2 \sqrt{2 \alpha  q^2+r^4}}dt + p \cos\theta d\phi\,,
\eea
where $F_1$ is the Appell hypergeometric function. The electromagnetic duality is manifest since the last two arguments in the Appell hypergeometric function are symmetric in exchange.
When $p=0$ or $q=0$, the Appell hypergeometric function collapses to the hypergeometric function, and the solution reduces respectively to the electric or magnetic black hole of Einstein-BI gravity, as can be expected. When $p=q$, the solution reduces to the usual dynonic RN black hole of equal electric and magnetic charges.

The dyonic black hole for our new self-dual theories \eqref{emdualf2} and \eqref{emdualf3} are much more complicated. The electric field of the \eqref{emdualf2} theory satisfies the algebraic relation
\be
\psi = \frac{q \sqrt{\sqrt{-16 \alpha  \psi ^2+\frac{16 \alpha  p^2}{r^4}+1}+1}}{r^2 \sqrt{3-\sqrt{-16 \alpha  \psi ^2+\frac{16 \alpha  p^2}{r^4}+1}}}\,.\label{newpsi}
\ee
When $p=q$, it has a simple solution $\psi = q/r^2$, and the metric becomes that of the dyonic RN black hole with equal electric and magnetic charges, as in the earlier simpler example. Although the algebraic equation \eqref{newpsi} can be solved explicitly, it is not worth presenting the result. We shall give only the large-$r$ expressions
\bea
\psi &=& \frac{q}{r^2}-\frac{4\alpha  q \left(q^2-p^2\right)}{r^6}-\frac{8\alpha ^2 q \left(p^4+2 p^2 q^2-3 q^4\right)}{r^{10}}\cr
&&+\frac{32 \alpha ^3 q \left(3 p^6-5 p^4 q^2+9 p^2 q^4-7 q^6\right)}{r^{14}} + \cdots\,,\nn\\
h &=& 1-\frac{2 M}{r}+\frac{p^2+q^2}{r^2}-\frac{2\alpha \left(p^2-q^2\right)^2}{5 r^6}+\frac{ 8\alpha ^2 \left(p^2-q^2\right)^2 \left(p^2+q^2\right)}{9 r^{10}}\cr
&& -\frac{8 \alpha ^3 \left(p^2-q^2\right)^2 \left(7 p^4+2 p^2 q^2+7 q^4\right)}{13 r^{14}} + \cdots\,.
\eea
For our other self-dual theory \eqref{emdualf3}, the purely electric and magnetic solution can be constructed explicitly, giving
\be
h=1 - \fft{2M}{r} + \frac{4}{5 \alpha  r} \left(5 r^3-\frac{\sqrt{2} \left(4 \alpha ^2 Q^4-20 \alpha  Q^2 r^4+5 r^8\right) \sqrt{\sqrt{2 \alpha  Q^2+r^4}+r^2}}{\left(r^4-2 \alpha  Q^2\right) \sqrt{2 \alpha  Q^2+r^4}-6 \alpha  Q^2 r^2+r^6}\right),
\ee
where $Q$ is either magnetic $p$ or electric $q$. An analytic solution of the general dyonic black hole is unlikely to exist, but the large $r$ expansion is given by
\bea
\psi &=& \fft{q}{r^2} + \frac{3 \alpha  q \left(p^2-q^2\right)}{4 r^6}-\frac{9 \alpha ^2 q \left(p^4+2 p^2 q^2-3 q^4\right)}{32 r^{10}}\nn\\
 &&+ \frac{\alpha ^3 q \left(35 p^6+3 p^4 q^2+105 p^2 q^4-143 q^6\right)}{128 r^{14}} + \cdots\,,\nn\\
h&=&1 - \fft{2M}{r} +\frac{p^2+q^2}{r^2} -\frac{3 \alpha  \left(p^2-q^2\right)^2}{40 r^6} + \frac{\alpha ^2 \left(p^2-q^2\right)^2 \left(p^2+q^2\right)}{32 r^{10}}\cr
&&-\frac{\alpha ^3 \left(p^2-q^2\right)^2 \left(143 p^4+146 p^2 q^2+143 q^4\right)}{6656 r^{14}} + \cdots\,.
\eea
In all these solutions, the electromagnetic duality $p\leftrightarrow q$ is manifest and the solution collapses to the dyonic RN black hole when $p=q$. It is also worth pointing out, as we have explained earlier, the black holes of all the self-dual theories are necessarily the same up to and including the $\alpha^2$ order, after matching the coupling constant appropriately. They diverge at the $\alpha^3$ order. While these dyonic black holes are certainly of interest, they are not regular, and are outside the scope of the main discussion of this paper. We have not yet managed to construct an exact regular dyonic black hole, which is in line with the observation that dyonic solutions are incompatible with a regular core \cite{Bronnikov:2022ofk}.

\section{Further properties}

\subsection{Repulson black holes and stars}

The special static metrics are completely determined by the equation
\be
rh' + h -1 = -\rho\,,
\ee
for a given matter energy density $\rho$. Thus regular solutions are necessarily fine-tuned by choosing the mass parameter precisely. Any deviation of this parameter with other parameters fixed will necessarily generate a singularity, naked or covered by an event horizon.
It is thus of interest to examine how the motions of a neutral particle are influenced by the regular solutions.

The solutions we have constructed all contain two integration constants $(M,p)$, together with the coupling constant $\alpha$ of the theory. They all have similar features and we therefore use the solution \eqref{oursol1} as an illustrative example.  Without loss of generality, we set $\alpha=1$, which has the effect of setting the scale of the system. The solution is regular when $M=M^{\rm cr}$. As illustrated in the right panel of Fig.~\ref{metrich}, when $p\ge 4.4873$, the metric is a black hole, the singularity must inevitably develop once any massive neutral particles fall into the black hole.

The situation becomes more interesting when $p<4.4873$, in which case, the metric describes a star with a dS core. Therefore, massive neutral particles will experience a repulsive gravity force. For this reason we call such a star a repulson. Note that gravity becomes attractive as usual once the particle is sufficiently far away from the core.

Repulsive force can also emerge in singular black holes, as shown in the right panel of Fig.~\ref{metrich}, when $M=1.1 M^{\rm cr}$ and $p=4$. The extra wiggle of the function $h$ outside the black hole implies that the gravity is repulsive in the region between the local maximum and minimum. Such black hole structures were also reported in dyonic black holes from Einstein gravity coupled to quasitopological electromagnetism \cite{Liu:2019rib}.

The geodesic motion of a massive ($\epsilon=1$) and massless ($\epsilon=0$) particle is governed by the conservation of the Hamiltonian
\be
H=\dot r^2 + h(r)\Big(\fft{L^2}{r^2} +\epsilon\Big) = E^2\,.
\ee
Defining $U= h(r)\Big(\fft{L^2}{r^2} +\epsilon\Big)-E^2$, a circular orbit at radius $r_0$ must satisfy
\be
U(r_0)=0=U'(r_0)\,.
\ee
The circular orbit is stable when $U''(r_0)>0$ and unstable when $U''(r_0)<0$.

\subsubsection{Photon spheres and shadows}

The circular orbits of photons ($\epsilon=0)$ are determined by $(h/r^2)'=0$, and their stability is determined by the sign of $(h/r^2)''$. Since $h(r\rightarrow\infty)\rightarrow 1$, the quantity $(h/r^2)'$ negatively approaches zero at the asymptotic infinity. For stars or black holes, we have
\be
\hbox{star:}\qquad \Big(\fft{h}{r^2}\Big)'\Big|_{r\rightarrow 0} <0\,;\qquad\qquad
\hbox{black hole:}\qquad \Big(\fft{h}{r^2}\Big)'\Big|_{r\rightarrow r_+} >0\,.
\ee
Thus a star can only have an even number of photon spheres, including zero. A black hole can only have an odd number of photon spheres, with at least one.  In either case, the outmost photon sphere is the unstable one and alternatingly stable and unstable between the two adjacent photon spheres.

We now examine the concrete example of the star solution given in section \ref{sec:simple}, with $M=M^{\rm cr}$. We can rescale all the lengthy quantities by a factor of $\sqrt{\alpha}$ so that they become dimensionless, and then set $\alpha=1$ for simplicity. The solution describes a star provided that $p<p^*=4.4873$. For sufficiently small $p$, there is no photon sphere. When $p\ge 4.02268$, two photon spheres emerge. The outer unstable photon sphere increases its size as $p$ increases. The inner stable photon sphere decreases in size until it touches the horizon of the extremal black hole that the star forms at $p=p^*$. For $p>p^*$, there exists only one unstable photon sphere that expands with the increasing of $p$.

The repulson black holes have two independent parameters $(M, p)$, and they are characterized by the extra wiggle of the metric function $h$ outside the horizon. We shall not be exhaustive in its analysis since a similar phenomenon occurred in dyonic black holes from Einstein gravity coupled to quasi-topological electromagnetism \cite{Liu:2019rib}. It should be pointed out that the wiggle does not guarantee that there must be an extra photon sphere. For example, for $M=1.1 M^{\rm cr}$, the $p=3$ solution is a repulson black hole, but with only one photon sphere at $r_1=0.728832$, outside the horizon $r_+=0.507009$. When $p=3.5$, the black hole with $r_+=0.659149$ remains a repulson, but now it has three photon spheres, with radii $r_1=0.920509$, $r_2=3.51253$ and $r_3=6.10713$.  Black holes with three photon spheres with the middle one being stable were first reported in \cite{Liu:2019rib}. More black holes with multiple photon spheres were later found in \cite{Gan:2021pwu,Guo:2022ghl}.

The outermost photon sphere will give rise to a black hole shadow of the radius
\be
r_{\rm sh} = \fft{r_{\rm ph}}{\sqrt{h(r_{\rm ph})}}\,.
\ee
It was conjectured that black hole parameters satisfy a sequence of inequalities \cite{Lu:2019zxb}
\be
\ft32 r_+\, \le\, r_{\rm ph}\, \le \,\fft{r_{\rm sh}}{\sqrt3} \,\le\, 3 M,\label{conjecture}
\ee
which include those obtained in \cite{Hod:2013jhd,Cvetic:2016bxi}. The relevant energy conditions were established in \cite{Yang:2019zcn}.  The black holes constructed from the $f(F^2)$ theory in this paper, singular or regular, all obey the above inequality. For the regular star solution that admits photon spheres, the above inequalities are also satisfied, although the leftmost inequality does not apply.

\subsubsection{Static massive shell, circular orbits}

We now consider circular orbits of massive particles. We shall focus on the repulson black holes or stars since our analysis would yield no interesting results for a typical black hole. For the star solution in question, at a given $p$, there is an $r_{\rm min}$ such that $h'(r_{\rm min})=0$. Gravity is attractive for $r<r_{\rm min}$ and hence there can be no circular orbits. Circular orbits can exist at all $r_0>r_{\rm min}$, and they are all stable for sufficiently small $p$.

Unstable circular orbits develop when $p$ is sufficiently large, but not large enough to form a black hole. Concretely, we consider $p=4$, we have $r_{\rm min}=3.9927$, $r_1=5.5511$ and $r_2=12.929$ such that
\be
U(r_i)=U'(r_i)=U''(r_i)=0\,,\qquad i=1,2.
\ee
circular orbits exist only outside the static massive shell at $r_{\rm min}$. We find
\bea
\hbox{Stable orbits:}&&\qquad r_0>r_2\,,\qquad r_{\rm min}<r_0<r_1\,,\nn\\
\hbox{Unstable orbits:}&&\qquad r_1<r_0<r_2\,.
\eea
We therefore refer $r_1$ to the ``local outermost stable circular orbit,'' or LOSCO, and refer $r_2$ to the ``local innermost stable circular orbit,'' or LISCO. These properties suggest that the repulson stars may be able to withstand the perturbation from forming a black hole.

We now examine the circular orbits of repulson black holes. We shall not be exhaustive in exploring the two parameters $(M,p)$.  Instead we consider a specific example, depicted in the left panel of Fig.~\ref{metrich}, i.e. $M=1.1 M^{\rm cr}$. Setting $\alpha=1$ again, the black hole develops three horizons when $p>3.66215$. Thus the repulson black hole with a single horizon exists when we have $0<p<3.66215$. We consider two specific cases, $p=3$ and $p=3.5$. Table 1
lists some essential data of these two black holes. Here, $r_+$ denotes the horizon radius, $r_{\rm ph}^i$ are the radius of the photon spheres, with $i=1,2,3$. The maximum and minimum of the metric function $h$ are located at $r_{\rm max}$ and $r_{\rm min}$ respectively, and $r_{\rm LISCO}$ is the LISCO location.

\bigskip
\begin{center}
\begin{tabular}{|c|c|c|c|c|c|c|c|}
  \hline
   $p$ & $r_+$ & $r_{\rm ph}^1$ & $r_{\rm max}$ & $r_{\rm min}$ & $r_{\rm ph}^2$ & $r_{\rm ph}^3$ & $r_{\rm LISCO}$ \\ \hline
  3.0 & 0.50701 & 0.72883 & 1.18561 & 2.99809 & NA & NA & NA \\ \hline
  3.5 & 0.65915 & 0.92051 & 1.2806 & 3.2372 & 3.51253 & 6.10713 & 12.000\\
  \hline
\end{tabular}
\bigskip

Table 1. {\it\small Relevant data of the two repulson black holes with $M=M^{\rm cr}$ and $\alpha=1$.}
\end{center}

With the data presented in Table 1, we can easily describe the properties of massive circular orbits. When $p=3$, circular orbits are not possible within the innermost photon sphere $r_{\rm ph}^1$, since it would require imaginary orbital angular momentum $L$. Circular orbits emerge in the region $(r_{\rm ph}^1, r_{\rm max})$, i.e. from the unstable photon sphere to $r_{\rm max}$, which is the location of the unstable static massive shell. No circular orbits could exist in $(r_{\rm max}, r_{\rm min})$ since gravity is repulsive there. A stable static massive shell can form at $r_{\rm min}$. We find that for all $r>r_{\rm min}$, there can be stable circular orbits.

The circular orbits have even richer properties when $p=3.5$.  The circular orbits in $(r_{\rm ph}^1, r_{\rm max})$, $(r_{\rm min}, r_{\rm ph}^2)$, $(r_{\rm ph}^3, r_{\rm LISCO})$, $(r_{\rm LISCO},\infty)$ are unstable, stable, unstable, stable, respectively.  In regions that are not mentioned above, there is no circular orbit since it would require imaginary values of $E$, $L$ or both.

\subsection{Cosmological constant from constant electric field}
\label{sec:constante}

We have so far focused on the spacetime structures that are asymptotic to Minkowski. It is straightforward to introduce a bare cosmological constant $\Lambda_0$ in the Lagrangian of Einstein-$f(F^2)$ theory. The metric remains specially static and the function $h$ is simply augmented by a term $-\Lambda_0 r^2/3$. It should be clarified that our second no-go theorem is no longer applicable, since one can always fine-tune the cosmological constant so that the dS core becomes Minkowski core.

In this subsection, we would like to point out an intriguing phenomenon that for certain $f(F^2)$ theories, a cosmological constant can be generated from a constant electric field provided that $\varphi (F^2)$ has zeros. (See also \cite{Bokulic:2022cyk}.) The effective cosmological constant is given by
\be
\Lambda_{\rm eff} = -\ft12 f(-2\psi_0^2)\,,\qquad \hbox{with}\qquad \varphi(-2\psi_0^2)=0\,.
\ee
As an example, we consider
\be
f=-\beta ( F^2 + \alpha (F^2)^2)\,.
\ee
In this case, for a purely electric configuration $F=\psi dr\wedge dt$, Maxwell's equation of motion can be solved with constant $\psi_0=1/(2\sqrt{\alpha})$ and hence the effective cosmological constant is generated:
\be
\Lambda_{\rm eff}=-\fft{\beta}{8\alpha}\,.
\ee
For general nonconstant $\psi$, it satisfies
\be
\beta \psi (1 - 4 \alpha \psi^2) = \fft{q}{r^2}\,.
\ee
Thus for $r$ runs from 0 to $\infty$, the electric field $\psi$ runs from $\infty$ to $\psi_0$ instead of zero, giving rise to an electrically charged black hole that is asymptotic to (A)dS.

Another interesting example is $f=-\beta F^2 \sqrt{1 + \alpha F^2}$. The Maxwell equation is
\be
\fft{\beta (1-3\alpha \psi^2)\psi}{\sqrt{1-2\alpha \psi^2}} = \fft{q}{r^2}\,.
\ee
Thus we have $\psi_0=1/\sqrt{3\alpha}$, giving rise to $\Lambda_{\rm eff}=-\beta/(3\sqrt3\,\alpha)$. In this case, the electric field $\psi$ runs from $\psi_1=1/\sqrt{2\alpha}$ to $\psi_0$ as $r$ runs from zero to infinity.

\section{Conclusions}
\label{sec:con}

We studied Einstein gravity minimally coupled to a particular type of NLED in the form of $f(F^2)$. The spherically-symmetric and static solution will necessarily take the special static metric \eqref{ssm}. We presented several classes of analytic functions of $f(F^2)$ and obtained explicit regular black holes or stars that satisfy the DEC. In the weak field limit, these $f(F^2)$ theories can be viewed as perturbations to Maxwell's theory. Our construction allows one to study both the theories and black holes further at a deeper level. Table 2 lists some major properties of various black holes.

\begin{center}
\begin{tabular}{|c|c|c|c|c|c|c|}
  \hline
   & WEC & DEC & SEC & TEC & Regular &$f$-Analytic\\
  RN & \checkmark & \checkmark & \checkmark & \checkmark & $\times$ &\checkmark \\
 BI & \checkmark & \checkmark & \checkmark & \checkmark & $\times$ &\checkmark \\
  Bardeen & \checkmark & $\times$ & $\times$ & $\times$ & \checkmark &$\times$\\
  Hayward & \checkmark & $\times$ & $\times$ & $\times$ & $\times$ &$\times$\\
  Ours & \checkmark & \checkmark & $\times$ & \checkmark & \checkmark &\checkmark\\
  \hline
\end{tabular}
\bigskip

Table 2: {\it \small Properties of various black holes in Einstein gravity coupled to NLED.}
\end{center}

We managed to prove that regular black holes and stars with special static metrics \eqref{ssm} must violate the SEC and such black holes with Minkowski core must violate the NEC. Such violation may not be surprising for the regular special static metrics since the time ticking rates of static objects sitting at asymptotic infinity and at the core are the same. This is certainly counterintuitive since we know that there is a gravitational time dilation in the middle of a real star such as the Sun. Therefore we should not overstate our conclusion to include the more general static metrics where $g_{tt}g_{rr}$ is not constant, in which case the situation remains far from being clear.

It should be emphasized again that regular black holes are fine-tuned objects where the free parameter mass in the general solution is taken to be some specific value to avoid the curvature singularity at the core. We therefore examined the geodesic motions of the regular solutions. When the charge $p$ is sufficiently small, the solution is a star with repulsive static gravity force at the core. We therefore call such a solution a repulson star. There is a static potential minimum where a massive static shell can accumulate. Gravity becomes attractive outside this shell, and circular orbits are stable if they are sufficiently close to the shell. These imply that the repulson stars are perturbatively stable from collapsing to a black hole. However, if $p$ is sufficiently large and the metric becomes a black hole, there is nothing to prevent it from attracting massive particles to develop singularity in the core.  For singular black holes, there can exist 1, 2 or 3 horizons. The single horizon black hole can also be repulson-like, and its perturbative stability requires further investigation.  Both repulson stars and black holes can attract multiple photon spheres with a stable one shielded by the shadow if observed from afar.

Our explicit examples of regular black holes carry magnetic charges, and $F^2$ diverges at the core. We developed a formalism by making use of the scalar-vector theory to perform electromagnetic duality transformation in $f(F^2)$. This allows us to obtain the dual theory of these regular black hole metrics so that they carry the electric charges instead. Intriguingly this cures the problem of the divergence of $F^2$ at the core in the magnetic cases. The $F^2$ is now finite from the core $(r=0)$ to asymptotic infinity ($r\rightarrow \infty$) for the dual electrically-charged regular black holes. We also applied formalism of the electromagnetic duality to obtain two new examples \eqref{emdualf2} and \eqref{emdualf3} of NLED that have electromagnetic duality. We studied the properties of their dyonic black holes that exhibit manifest electromagnetic duality.

Our final observation is that an effective cosmological constant can emerge from a constant electric field in some appropriate $f(F^2)$. However, the homogeneity and isotropicity of the space are broken by the tensor field strength, even though the metric is homogeneous. Therefore, such NLED should belong to the dark sector that interacts with normal matter only through gravitation. It is of interest to investigate their physical implication in cosmology.

\section*{Acknowledgement}

We are grateful to Gary Gibbons and Run-qiu Yang for useful discussions. This work is supported in part by NSFC (National Natural Science Foundation of China) Grants No.~11935009 and No.~11875200.

\end{document}